\documentclass[letterpaper,12pt]{article}   
\usepackage{osajnl2} 
\usepackage[draft]{hyperref} 
\usepackage{amsmath}
\usepackage{amssymb}

\begin{document}

\title{Preparation and purification of four-photon Greenberger-Horne-Zeilinger state}


\author{Ying-Qiu He,$^{1}$ Dong Ding, $^{1,2}$ Feng-Li Yan $^{1,4}$ and Ting Gao$^{3,5}$}
\address{$^1$College of Physics Science and Information Engineering, Hebei Normal University, Shijiazhuang 050024, China}
\address{$^2$Department of Basic Curriculum, North China Institute of Science and Technology, Beijing 101601, China}
\address{$^3$College of Mathematics and Information Science, Hebei Normal University, Shijiazhuang 050024, China}
\address{$^4$ flyan@hebtu.edu.cn}
\address{$^5$ gaoting@hebtu.edu.cn}

\begin{abstract}
We present an efficient scheme for preparing and purifying of four-photon Greenberger-Horne-Zeilinger (GHZ) state based on linear optics and postselection.
First, we describe how to create a four-photon GHZ state in both polarization and spatial degrees of freedom from two pairs. Moreover, in the presence of depolarization noise our scheme is capable of purifying the desired state. In the regime of weak nonlinearity we design an indirect photon number-resolving detection to distinguish two states of the two pairs. At last, a fourfold coincidence detector click indicates creating of a polarization-entangled four-photon GHZ state.
\end{abstract}


\maketitle 

\section{Introduction}
Multipartite entanglement \cite{HHHH2009, GT2009} is one of the most fundamental and intriguing features of quantum mechanics and a resource for a variety of quantum information processing tasks \cite{NC2000, GHZ1990, W2000, FD2011, Yan2011, GYE2014}. Currently, optical quantum systems \cite{Kok2007, Pan2012} are prominent candidates for quantum information processing. Based on the  inherent nature of photons,  a photon represents a qubit and photons can be made to interact with each other using cross-Kerr nonlinearities \cite{Milburn1984, Imoto1985}. Remarkably, in 2001, Knill \emph{et al} \cite{KLM2001} have shown a scalable quantum computation, in principle, with linear optics, single-photon sources and detectors. Motivated by this report, a series of searches for optical quantum information processing have been proposed in succession \cite{SimonPan2002, ShengDeng2010, DYG2013, DingYanGao2012}.

There are several methods to prepare multi-photon entangled states based on linear optics and cross-Kerr nonlinearities.
As a standard method, experimentally, a parametric down-conversion (PDC) source can emit one pair of polarization-entangled photons with a given probability \cite{Kwiat1995}.
Then, one can transform two pairs of polarization entangled photons into three-photon Greenberger-Horne-Zeilinger (GHZ) state \cite{GHZExperiment99}, four-photon GHZ state \cite{Pan2001}, hyperentangled state \cite{DuQiao2012, DingYan2013}, and so on.
On the other hand, in the regime of  weak cross-Kerr nonlinearities, one can construct quantum nondemolition detection (QND) \cite{GLP1998, MNBS2005, NM2004, Barrett2005, Kok2008, HRB2009, LHBR2009} to project signal photons onto a desired subspace. In general, the signal modes may be fed by several single photons or some entangled photons. For example, Nemoto and Munro \cite{NM2004} proposed a two polarization qubit entangling gate, and Barrett \emph{et al} \cite{Barrett2005} described a method of nondestructive Bell-state detection, where the detections of coherent probe beams in two schemes are all homodyne measurement of the position quadrature.

In this paper, we focus on the preparation and purification of four-photon GHZ state. Especially, we describe a method to purify the desired four-photon state in depolarization-noise channel.

\section{Creation of four-photon eight-qubit Greenberger-Horne-Zeilinger state}
As is well known, when a short pulse of ultraviolet light passes through the $\beta$-barium-borate (BBO) crystal, a pair of polarization entangled photons may be prepared in spatial modes $a_1$ and $b_1$, as shown in Fig.1. Also, another pair may be emitted in spatial modes $a_2$ and $b_2$ because of the function of a mirror. We suppose that two independent photon pairs are created in state
\begin{equation}\label{product state}
\frac{1}
{2}\left( {{{\left| H \right\rangle }_{a_i}}{{\left| V \right\rangle }_{b_i}} - {{\left| V \right\rangle }_{a_i}}
{{\left| H \right\rangle }_{b_i}}} \right)\left( {{{\left| H \right\rangle }_{a_j}}{{\left| V \right\rangle }_{b_j}}
- {{\left| V \right\rangle }_{a_j}}{{\left| H \right\rangle }_{b_j}}} \right),
\end{equation}
where $a_{i,j}$ and $b_{i,j}$ ($i,j=1,2$) respectively represent four spatial modes, $\left| H \right\rangle$ ($\left| V \right\rangle$) indicates the state of a horizontally (vertically) polarized photon. Obviously, there are two cases, that is, $i=j$ and $i \neq j$.
For $i=j$, the two pairs are superposition of two spatial modes $a_1$ and $b_1$ and the other two $a_2$ and $b_2$. While, for $i \neq j$, it means that there is one photon in each spatial modes $a_1$, $b_1$, $a_2$ and $b_2$, respectively.

In Fig.1, each polarizing beam splitter (PBS) is used to transmit $|H\rangle$ polarization photons and reflect $|V\rangle$ polarization photons.
The beam splitters (BSs) represent $50:50$ polarization-independent beam splitters. The half wave plate $ \texttt{R}_{90}$ is used to convert the polarization state $|H\rangle$ into $|V\rangle$ or vice versa. $C_1$, $c_1$, $C_2$, $c_2$, $D_i$, $d_i$, $E_i$, and $e_i$ ($i=1,2,3,4$) are the different spatial modes, respectively.
Through the setup towards the spatial modes $D_i$ and $d_i$ the involved components of Eq.(\ref{product state}) evolve as
\begin{equation}\label{}
|H\rangle_{a_{1}} \rightarrow \frac{1} {{\sqrt 2 }}\left(\left| H \right\rangle_{D_{1}} + \left| H \right\rangle_{D_{3}}\right),~~~
|H\rangle_{a_{2}} \rightarrow \frac{1} {{\sqrt 2 }}\left(\left| H \right\rangle_{d_{1}} + \left| H \right\rangle_{d_{3}}\right),
\end{equation}
\begin{equation}\label{}
|V\rangle_{a_{1}} \rightarrow \frac{1} {{\sqrt 2 }}\left(\left| V \right\rangle_{D_{1}} + \left| V \right\rangle_{D_{2}}\right),~~~
|V\rangle_{a_{2}} \rightarrow \frac{1} {{\sqrt 2 }}\left(\left| V \right\rangle_{d_{1}} + \left| V \right\rangle_{d_{2}}\right),
\end{equation}
\begin{equation}\label{}
|H\rangle_{b_{1}} \rightarrow \frac{1} {{\sqrt 2 }}\left(\left| H \right\rangle_{D_{2}} + \left| H \right\rangle_{D_{4}}\right),~~~
|H\rangle_{b_{2}} \rightarrow \frac{1} {{\sqrt 2 }}\left(\left| H \right\rangle_{d_{2}} + \left| H \right\rangle_{d_{4}}\right),
\end{equation}
\begin{equation}\label{}
|V\rangle_{b_{1}} \rightarrow \frac{1} {{\sqrt 2 }}\left(\left| V \right\rangle_{D_{3}} + \left| V \right\rangle_{D_{4}}\right),~~~
|V\rangle_{b_{2}} \rightarrow \frac{1} {{\sqrt 2 }}\left(\left| V \right\rangle_{d_{3}} + \left| V \right\rangle_{d_{4}}\right).
\end{equation}

Consider now the case that the four photons are emitted, one in each spatial mode $D_1$ (or $d_1$), $D_2$ (or $d_2$), $D_3$ (or $d_3$), and $D_4$ (or $d_4$).
It can be designed to obtain the desired states by a set of fourfold coincidence detections associated with spatial mode $E_i$ and $e_i$ ($i=1,2,3,4$).
Specifically, for $i=j$, after the equipment the satisfactory state reads
\begin{equation}\label{}
{\left| {\phi _{0}} \right\rangle} =\frac{1}{2}\left(\left| HVVH \right\rangle+\left| VHHV \right\rangle\right)\otimes\left(\left|{D_1D_2D_3D_4}\right\rangle+\left|{d_1d_2d_3d_4}\right\rangle \right).\end{equation}
While for $i \neq j$, the state evolves as \cite{DuQiao2012, DingYan2013}
\begin{eqnarray}
{\left| {\psi _{0}} \right\rangle}& = &\frac{1}{2}\left(\left| H \right\rangle_{D_1}\left| V \right\rangle_{d_2}\left| V \right\rangle_{D_3}\left| H \right\rangle_{d_4}
+\left| H \right\rangle_{d_1}\left| V \right\rangle_{D_2}\left| V \right\rangle_{d_3}\left| H \right\rangle_{D_4}\right) \nonumber \\
& & +\frac{1}{2}\left(\left| V\right\rangle_{d_1}\left| H \right\rangle_{d_2}\left| H \right\rangle_{D_3}\left| V \right\rangle_{D_4}
+\left| V \right\rangle_{D_1}\left| H \right\rangle_{D_2}\left| H \right\rangle_{d_3}\left| V \right\rangle_{d_4}
\right).
\end{eqnarray}
Then, after four half wave plates and PBSs, one obtains the desired four-photon entangled state containing entanglement in both polarization and spatial degrees of freedom. That is, for $i=j$, the state
\begin{equation}\label{ }
{\left| {\phi _{}} \right\rangle} =\frac{1}{2}\left(\left| HVVH \right\rangle+\left| VHHV \right\rangle\right) \left(\left|{e_1E_2E_3e_4}\right\rangle+\left|{E_1e_2e_3E_4}\right\rangle \right)\end{equation}
can be created. While for $i \neq j$, we obtain the state
\begin{equation}\label{ }
{\left| {\psi} \right\rangle} =\frac{1}{2}\left[\left(\left| HHVV \right\rangle+\left| VVHH \right\rangle\right) \left|{e_1E_2E_3e_4}\right\rangle
         + \left(\left| HVHV\right\rangle+\left| VHVH \right\rangle\right) \left|{E_1e_2e_3E_4}\right\rangle\right].
\end{equation}

By far, we describe an efficient scheme to create four-photon eight-qubit GHZ states which are entangled in polarization and spatial modes. We derive two representations, which contain all possible cases about emitted four photons, namely, $i=j$ and $i \neq j$.

\section{Polarization entanglement purification using spatial entanglement}
In practice, we note that the prepared entangled photons may suffer from channel noise when traveling from a source to a destination (i.e., $d_i$ and $D_i$ ($i=1,2,3,4$)). Because an impact of noise on spatial modes can be easily avoided \cite{SimonPan2002}, we here suppose that only the polarization entanglement suffers from channel noise \cite{ShengDeng2010, DingYan2013}. The following task of our scheme is to purify the polarization entanglement by using spatial entanglement based on linear optics and multi-fold coincidence detections.

In a depolarization channel, the created entangled states ${\left| {\phi _{0}} \right\rangle}$ and ${\left| {\psi _{0}} \right\rangle}$ may suffer from bit-flip and phase-flip errors. Now, let us take the state ${\left| {\psi _{0}} \right\rangle}$ for example. We here suppose that two spatial modes $d_1$ and $D_1$ are faultless and only the photons in spatial modes $d_i$ and $D_i$ ($i=2,3,4$) suffer depolarization.
Clearly, the initial state may be transformed to one of the following sixteen four-photon entangled states:
\begin{equation}\label{}
{\left| {\psi _{1}^ \pm } \right\rangle} = \frac{1}
{2}\left[ {\left| {HVVH} \right\rangle \left( {{d_1}{D_2}{d_3}{D_4} + {D_1}{d_2}{D_3{d_4}}} \right) \pm \left| {VHHV} \right\rangle \left( {{D_1}{D_2}{d_3}{d_4} + {d_1}{d_2}{D_3}{D_4}} \right)} \right],
\end{equation}
\begin{equation}\label{}
{\left| {\psi _{2}^ \pm } \right\rangle} = \frac{1}
{2}\left[ {\left| {HVHV} \right\rangle \left( {{d_1}{D_2}{d_3}{D_4} + {D_1}{d_2}{D_3{d_4}}} \right) \pm \left| {VHVH} \right\rangle \left( {{D_1}{D_2}{d_3}{d_4} + {d_1}{d_2}{D_3}{D_4}} \right)} \right],
\end{equation}
\begin{equation}\label{}
{\left| {\psi _{3}^ \pm } \right\rangle} = \frac{1}
{2}\left[ {\left| {HHVV} \right\rangle \left( {{d_1}{D_2}{d_3}{D_4} + {D_1}{d_2}{D_3{d_4}}} \right) \pm \left| {VVHH} \right\rangle \left( {{D_1}{D_2}{d_3}{d_4} + {d_1}{d_2}{D_3}{D_4}} \right)} \right],
\end{equation}
\begin{equation}\label{}
{\left| {\psi _{4}^ \pm } \right\rangle} = \frac{1}
{2}\left[ {\left| {HVVV} \right\rangle \left( {{d_1}{D_2}{d_3}{D_4} + {D_1}{d_2}{D_3{d_4}}} \right) \pm \left| {VHHH} \right\rangle \left( {{D_1}{D_2}{d_3}{d_4} + {d_1}{d_2}{D_3}{D_4}} \right)} \right],
\end{equation}
\begin{equation}\label{}
{\left| {\psi _{5}^ \pm } \right\rangle} = \frac{1}
{2}\left[ {\left| {HVHH} \right\rangle \left( {{d_1}{D_2}{d_3}{D_4} + {D_1}{d_2}{D_3{d_4}}} \right) \pm \left| {VHVV} \right\rangle \left( {{D_1}{D_2}{d_3}{d_4} + {d_1}{d_2}{D_3}{D_4}} \right)} \right],
\end{equation}
\begin{equation}\label{}
{\left| {\psi _{6}^ \pm } \right\rangle} = \frac{1}
{2}\left[ {\left| {HHVH} \right\rangle \left( {{d_1}{D_2}{d_3}{D_4} + {D_1}{d_2}{D_3{d_4}}} \right) \pm \left| {VVHV} \right\rangle \left( {{D_1}{D_2}{d_3}{d_4} + {d_1}{d_2}{D_3}{D_4}} \right)} \right],
\end{equation}
\begin{equation}\label{}
{\left| {\psi _{7}^ \pm } \right\rangle} = \frac{1}
{2}\left[ {\left| {HHHV} \right\rangle \left( {{d_1}{D_2}{d_3}{D_4} + {D_1}{d_2}{D_3{d_4}}} \right) \pm \left| {VVVH} \right\rangle \left( {{D_1}{D_2}{d_3}{d_4} + {d_1}{d_2}{D_3}{D_4}} \right)} \right],
\end{equation}
\begin{equation}\label{}
{\left| {\psi _{8}^ \pm } \right\rangle} = \frac{1}
{2}\left[ {\left| {HHHH} \right\rangle \left( {{d_1}{D_2}{d_3}{D_4} + {D_1}{d_2}{D_3{d_4}}} \right) \pm \left| {VVVV} \right\rangle \left( {{D_1}{D_2}{d_3}{d_4} + {d_1}{d_2}{D_3}{D_4}} \right)} \right].
\end{equation}
After the action of the four half wave plates and PBSs, as shown in Fig.1, the combined system then evolves as
\begin{equation}\label{}
{\left| {{\psi_{1}^ \pm }} \right\rangle} \to \frac{1}
{2}\left[ {\left( {\left| {HHVV} \right\rangle  + \left| {VVHH} \right\rangle } \right)\left| {{e_1}{E_2}{E_3}{e_4}} \right\rangle  \pm \left( {\left| {HVHV} \right\rangle  + \left| {VHVH} \right\rangle } \right)\left| {{E_1}{e_2}{e_3}{E_4}} \right\rangle } \right],
\end{equation}
\begin{equation}\label{}
{\left| {{\psi_{2}^ \pm }} \right\rangle} \to \frac{1}
{2}\left[ {\left( {\left| {HHHH} \right\rangle  + \left| {VVVV} \right\rangle } \right)\left| {{e_1}{E_2}{e_3}{E_4}} \right\rangle  \pm \left( {\left| {HVVH} \right\rangle  + \left| {VHHV} \right\rangle } \right)\left| {{E_1}{e_2}{E_3}{e_4}} \right\rangle } \right],
\end{equation}
\begin{equation}\label{}
{\left| {{\psi_{3}^ \pm }} \right\rangle} \to \frac{1}
{2}\left[ {\left( {\left| {HVVH} \right\rangle  + \left| {VHHV} \right\rangle } \right)\left| {{e_1}{e_2}{E_3}{E_4}} \right\rangle  \pm \left( {\left| {HHHH} \right\rangle  + \left| {VVVV} \right\rangle } \right)\left| {{E_1}{E_2}{e_3}{e_4}} \right\rangle } \right],
\end{equation}
\begin{equation}\label{}
{\left| {{\psi_{4}^ \pm }} \right\rangle} \to \frac{1}
{2}\left[ {\left( {\left| {HHVH} \right\rangle  + \left| {VVHV} \right\rangle } \right)\left| {{e_1}{E_2}{E_3}{E_4}} \right\rangle  \pm \left( {\left| {HVHH} \right\rangle  + \left| {VHVV} \right\rangle } \right)\left| {{E_1}{e_2}{e_3}{e_4}} \right\rangle } \right],
\end{equation}
\begin{equation}\label{}
{\left| {{\psi_{5}^ \pm }} \right\rangle} \to \frac{1}
{2}\left[ {\left( {\left| {HHHV} \right\rangle  + \left| {VVVH} \right\rangle } \right)\left| {{e_1}{E_2}{e_3}{e_4}} \right\rangle  \pm \left( {\left| {HVVV} \right\rangle  + \left| {VHHH} \right\rangle } \right)\left| {{E_1}{e_2}{E_3}{E_4}} \right\rangle } \right],
\end{equation}
\begin{equation}\label{}
{\left| {{\psi_{6}^ \pm }} \right\rangle} \to \frac{1}
{2}\left[ {\left( {\left| {HVVV} \right\rangle  + \left| {VHHH} \right\rangle } \right)\left| {{e_1}{e_2}{E_3}{e_4}} \right\rangle  \pm \left( {\left| {HHHV} \right\rangle  + \left| {VVVH} \right\rangle } \right)\left| {{E_1}{E_2}{e_3}{E_4}} \right\rangle } \right],
\end{equation}
\begin{equation}\label{}
{\left| {{\psi_{7}^ \pm }} \right\rangle} \to \frac{1}
{2}\left[ {\left( {\left| {HVHH} \right\rangle  + \left| {VHVV} \right\rangle } \right)\left| {{e_1}{e_2}{e_3}{E_4}} \right\rangle  \pm \left( {\left| {HHVH} \right\rangle  + \left| {VVHV} \right\rangle } \right)\left| {{E_1}{E_2}{E_3}{e_4}} \right\rangle } \right],
\end{equation}
\begin{equation}\label{}
{\left| {{\psi_{8}^ \pm }} \right\rangle} \to \frac{1}
{2}\left[ {\left( {\left| {HVHV} \right\rangle  + \left| {VHVH} \right\rangle } \right)\left| {{e_1}{e_2}{e_3}{e_4}} \right\rangle  \pm \left( {\left| {HHVV} \right\rangle  + \left| {VVHH} \right\rangle } \right)\left| {{E_1}{E_2}{E_3}{E_4}} \right\rangle } \right].
\end{equation}

So we can conclude that the errors both from phase-flip and bit-flip can be corrected based on fourfold coincidence detection and postselection. More specifically, after evolution of the equipment, if one of the four-fold coincidence detections $e_1$, $e_2$, $E_3$, and $e_4$ (or $E_1$, $E_2$, $e_3$, and $E_4$) clicks, for example, we may conclude that the initial state ${\left| {\psi _{0}^{}} \right\rangle}$ suffers both phase-flip and bit-flip on one of the polarization qubits and evolves as ${\left| {\psi _{6}^{-}} \right\rangle}$. On the other hands, with the fourfold coincidence detector click, the four-photon hyperentangled state collapses into polarization entangle state $\frac{1}{\sqrt{2}}\left( {\left| {HVVV} \right\rangle  + \left| {VHHH} \right\rangle } \right)$ (or $\frac{1}{\sqrt{2}}\left( {\left| {HHHV} \right\rangle  + \left| {VVVH} \right\rangle } \right)$). In all, the phase-flip error can be erased in postselection and the bit-flip can be corrected by the following local bit-flip operation, which can be easily realized by some half wave plates. Furthermore, extending the same procedure allows us to analyze the case $i=j$.
It is worth noting that in order to obtain four-photon GHZ states rather than two-GHZ-pair states (entangled state $\frac{1}{2}\left( {\left| {HHVV} \right\rangle  + \left| {VVHH} \right\rangle } + {\left| {HVHV} \right\rangle  + \left| {VHVH} \right\rangle } \right)$, for example) there is an additional constraint (a set of spatial modes $d_1$ and $D_1$, for example, are faultless) for $i \neq j$, while for $i=j$ there is not the above constraint.
At last, without loss of generality, in order to obtain the four-photon polarization-entanglement GHZ state $\frac{1}{\sqrt{2}}\left( {\left| {HHHH} \right\rangle  + \left| {VVVV} \right\rangle } \right)$, for example, we list the required operations on the four photons corresponding to the results of fourfold coincidence detections, as shown in Table 1.

\begin{table}[h]
{\bf \caption{Operations on the four photons corresponding to the results of fourfold coincidence detections (FCD) for the cases $i=j$ and $i \neq j$, where $\hat I$ and ${\hat \sigma _x}$ represent the identity operator and bit-flip operation, respectively.}}\begin{center}
\begin{tabular}{ccc} \hline
 FCD  &  $i=j$   &  $i \neq j$ \\  \hline
${e_1}{e_2}{e_3}{e_4}$ & $\hat I \otimes \hat I  \otimes \hat I  \otimes \hat I $ &  $\hat I \otimes {\hat \sigma _x} \otimes \hat I  \otimes{\hat \sigma _x}$\\
${E_1}{E_2}{E_3}{E_4}$ & $\hat I \otimes \hat I  \otimes \hat I  \otimes \hat I $ &  $\hat I \otimes \hat I \otimes {\hat \sigma _x}  \otimes{\hat \sigma _x}$\\
${e_1}{e_2}{e_3}{E_4}$ & $\hat I \otimes \hat I  \otimes \hat I  \otimes{\hat \sigma _x}$ &  $\hat I \otimes{\hat \sigma _x} \otimes \hat I \otimes \hat I $\\
${E_1}{E_2}{E_3}{e_4}$ & $\hat I \otimes \hat I  \otimes \hat I  \otimes{\hat \sigma _x}$ &  $\hat I \otimes \hat I  \otimes {\hat \sigma _x} \otimes \hat I  $\\
${e_1}{e_2}{E_3}{e_4}$ & $\hat I \otimes \hat I  \otimes{\hat \sigma _x} \otimes \hat I $ &  $ {\hat \sigma _x} \otimes \hat I  \otimes \hat I  \otimes\hat I $\\
${E_1}{E_2}{e_3}{E_4}$ & $\hat I \otimes \hat I  \otimes{\hat \sigma _x} \otimes \hat I $ &  $\hat I \otimes \hat I  \otimes \hat I  \otimes{\hat \sigma _x}$\\
${e_1}{E_2}{e_3}{e_4}$ & $\hat I \otimes{\hat \sigma _x}  \otimes \hat I  \otimes \hat I $ & $\hat I \otimes \hat I \otimes \hat I  \otimes{\hat \sigma _x}$\\
${E_1}{e_2}{E_3}{E_4}$ & $\hat I \otimes{\hat \sigma _x}  \otimes \hat I  \otimes \hat I $ & ${\hat \sigma _x} \otimes \hat I \otimes \hat I  \otimes \hat I$\\
${E_1}{e_2}{e_3}{e_4}$ & ${\hat \sigma _x} \otimes \hat I  \otimes \hat I  \otimes \hat I $ & $\hat I \otimes {\hat \sigma _x} \otimes \hat I  \otimes \hat I $\\
${e_1}{E_2}{E_3}{E_4}$ & ${\hat \sigma _x} \otimes \hat I  \otimes \hat I  \otimes \hat I $ & $\hat I \otimes \hat I \otimes{\hat \sigma _x}\otimes \hat I $\\
${e_1}{e_2}{E_3}{E_4}$ & $\hat I \otimes \hat I\otimes{\hat \sigma _x}\otimes{\hat \sigma _x}$ &$\hat I \otimes {\hat \sigma _x}\otimes{\hat \sigma _x}\otimes \hat I $\\
${E_1}{E_2}{e_3}{e_4}$ & $\hat I \otimes \hat I  \otimes{\hat \sigma _x}\otimes{\hat \sigma _x} $ &  $\hat I \otimes \hat I  \otimes \hat I  \otimes \hat I$\\
${e_1}{E_2}{e_3}{E_4}$ & $\hat I \otimes{\hat \sigma _x} \otimes \hat I  \otimes{\hat \sigma _x}$ &  $\hat I \otimes \hat I  \otimes \hat I  \otimes \hat I$\\
${E_1}{e_2}{E_3}{e_4}$ & $\hat I \otimes{\hat \sigma _x} \otimes \hat I \otimes{\hat\sigma _x}$&$\hat I \otimes{\hat \sigma _x}\otimes{\hat \sigma _x} \otimes \hat I $\\
${e_1}{E_2}{E_3}{e_4}$ & $\hat I \otimes {\hat \sigma _x}\otimes{\hat\sigma _x}\otimes \hat I$ & $\hat I\otimes \hat I \otimes {\hat \sigma _x}\otimes{\hat \sigma _x}$\\
${E_1}{e_2}{e_3}{E_4}$ & $\hat I \otimes {\hat \sigma _x}\otimes{\hat\sigma _x} \otimes \hat I$ & $\hat I\otimes {\hat \sigma _x}\otimes \hat I\otimes{\hat \sigma _x}$\\
\hline
\end{tabular}
\end{center}
\end{table}
\clearpage

\section{Photon number-resolving detection}
As a matter of fact, if we cannot distinguish between $i=j$ and $i \neq j$, two cases of the created four-photon, when fourfold coincidence detector clicks we do not tell exactly what operation is required and what state has been prepared. However, it is noting that the action to distinguish between two cases must not destroy the nature of entangled photons. Thus, it becomes very important to distinguish the two cases in a nondestructive way.

Considering for $i=j$ there are two photons in each spatial mode $a_1$ and $b_1$ (or $a_2$ and $b_2$), while for $i \neq j$ there is one photon in each spatial mode $a_1$, $b_1$, $a_2$ and $b_2$, one can distinguish between two cases by deciding whether there is one photon in each spatial mode $a_1$ and $a_2$ or not. Now, in the regime of weak nonlinearities ($\theta\simeq10^{-2}$), we design a setup of QND to determine the four photons created by PDC sources whether in the case $i=j$ or in the alternative $i \neq j$. As shown in Fig.2, signal modes $a_1$ and $a_2$ are two spatial modes of the emitted photons, coherent probe beam $|\sqrt{2}\alpha\rangle$ followed by beam splitters, two single-qubit phase gates $R(-\theta)$s and cross-Kerr nonlinearities are used to construct cross-phase modulation, 1, 2, 3, and 4 are four spatial modes of the probe beams. After a series of transformations of the setup, the states of the combined system of the two signal modes and the probe beams
\begin{equation}\label{}
|1,1\rangle_{a_1a_2}|\alpha\rangle|\alpha\rangle
\end{equation}
and
\begin{equation}\label{}
\frac{1}{\sqrt{2}}(|0,2\rangle_{a_1a_2}+|2,0\rangle_{a_1a_2})|\alpha\rangle|\alpha\rangle,
\end{equation}
evolve as
\begin{equation}\label{}
|1,1\rangle_{a_1a_2}|\sqrt{2}\alpha\rangle_1 |0\rangle_2
\end{equation}
and
\begin{equation}\label{}
\frac{1}{\sqrt {2}}|\sqrt{2}\alpha \cos \theta\rangle_1 (|0,2\rangle_{a_1a_2}|\texttt{i}\sqrt{2}\alpha \sin \theta\rangle_2+|2,0\rangle_{a_1a_2}|-\texttt{i}\sqrt{2}\alpha \sin \theta\rangle_2),
\end{equation}
respectively. Here $|m,n\rangle_{a_1a_2}$ represents a two-spatial mode state, with $m$ photons in mode $a_1$ and $n$ photons in mode $a_2$. Then we introduce an indirect photon number-resolving detection \cite{HRB2009} on spatial mode 2 by using another coherent state $|\sqrt{2}\beta\rangle$. Here, the photon number-resolving detection serves two purposes. It is used to distinguish between the states $|0\rangle_2$ and $|\pm\texttt{i}\sqrt{2}\alpha \sin \theta\rangle_2$ but do not distinguish $|\texttt{i}\sqrt{2}\alpha \sin \theta\rangle_2$ from $|-\texttt{i}\sqrt{2}\alpha \sin \theta\rangle_2$ and project two signal photons onto a desired subspace, simultaneously.
After the evolution of BSs and cross-Kerr medium followed by a projection $|n\rangle\langle n|$ on spatial mode 4, one can obtain specific value of photon number $n$ in mode 2, as described by Lin \emph{et al} in \cite{LHBR2009}.
As a result, the value of projection $n=0$ corresponding to a vacuum state project two signal photons onto the state $|1,1\rangle_{a_1a_2}$ and then conclude that the two pairs are in case $i \neq j$. On the other hand, for $n\neq0$, two signal photons state is $\frac{1}{\sqrt{2}}(|0,2\rangle+|2,0\rangle)_{a_1a_2}$ up to a phase shift $\frac{n}{2}\pi$, which can be erased according to the classical feed-forward result $n$ of the projection $|n\rangle\langle n|$, and then claim that the two pairs are in case $i=j$.

Notice that interference occur, in principle, as two photons coming from two different pairs arrive at a polarizing beam splitter.
For the first experimental scheme of four-photon GHZ state \cite{Pan2001} or the preparation of four-photon twelve-qubit hyperentangled state \cite{DuQiao2012}, only fourfold coincidence detections or QND are used and thus the cases $i=j$ are discarded. Surprisingly, in our scheme the two cases are all collected, and so, in a sense, the present scheme is efficient.

\section{Discussion and summary}
In summary, we demonstrate a method of preparing and purifying four-photon GHZ state. The present scheme has several notable features.
Firstly, because it contains two cases, that is, two entangled photon pairs emitted in four spatial modes and alternatively only in two spatial modes, our scheme is more efficient than the conventional schemes.
Secondly, we design polarization entanglement purification using spatial entanglement and thus there is a powerful error-correcting capability in noisy channels.
Thirdly, by putting an H operation (a $45^\circ$ polarizer) followed by single-photon detectors in the $H/V$ basis in any one set of detectors ($e_4$ and $E_4$, for example), one can also implement preparation for three-photon GHZ state.
Finally, the strength of the nonlinearities required for the process of photon number-resolving detection are orders of magnitude weaker than threshold of becoming practical with electromagnetically induced transparency (i.e. $\theta\sim10^{-2}$).
In a word, we present an efficient preparation and purification scheme for four-photon GHZ state with current technology.
Also, the present scheme can be easily extended to preparing some other multi-photon entangled states and some other degrees of freedom, frequency degree of freedom, for example.

\section*{Acknowledgements}
This work was supported by the National Natural Science Foundation of China under Grant No: 11371005, Hebei Natural Science Foundation of China under Grant Nos: A2012205013, A2014205060,  the Fundamental Research Funds for the Central Universities of Ministry of Education of China under Grant No:3142014068, Langfang Key Technology Research and Development Program of China under Grant No: 2014011002.

\clearpage
\section*{List of Figure Captions}
Fig. 1. The schematic diagram of preparation and purification for four-photon GHZ state.

Fig. 2. A nondestructive photon number-resolving detection to distinguish between two entangled photon pairs emitted in four spatial modes (i.e. $i \neq j$) and only in two spatial modes (i.e. $i=j$).

\clearpage

\begin{figure}
  \centering\includegraphics[width=5in]{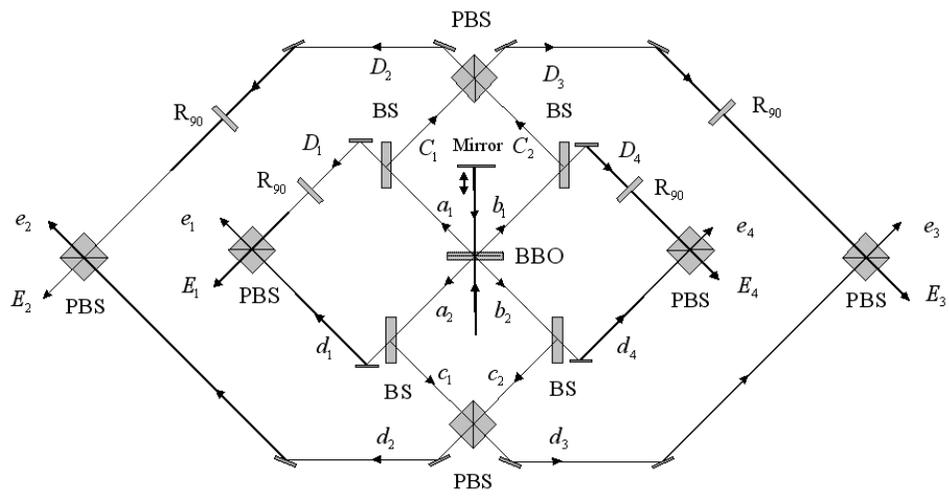}\\
  \caption{(color online). The schematic diagram of preparation and purification for four-photon GHZ state.}\label{}
\end{figure}

\clearpage
\begin{figure}
  \centering\includegraphics[width=4in]{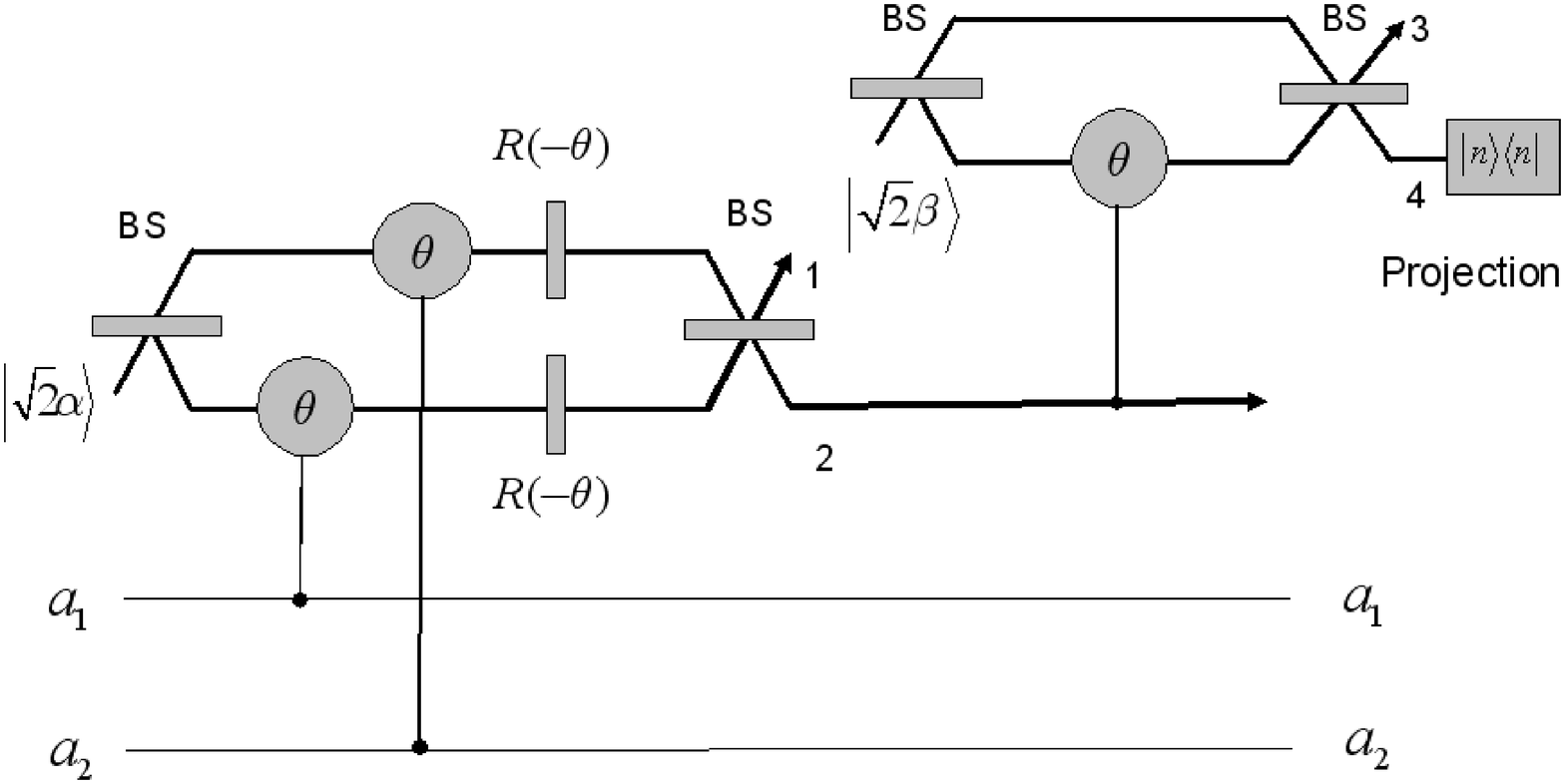}\\
  \caption{(color online). A nondestructive photon number-resolving detection to distinguish between two entangled photon pairs emitted in four spatial modes (i.e. $i \neq j$) and only in two spatial modes (i.e. $i=j$).}\label{}
\end{figure}


\begin{thebibliography}{99}
\bibitem{HHHH2009} R. Horodecki, P. Horodecki, M. Horodecki, and K. Horodecki, ``Quantum entanglement,'' Rev. Mod. Phys. \textbf{81}, 865--942 (2009).
\bibitem{GT2009} O. G$\ddot{\text{u}}$hne and G. T$\acute{\text{o}}$th, ``Entanglement detection,'' Phys. Rep. \textbf{474}, 1--75 (2009).
\bibitem{NC2000} M. A. Nielsen and I. L. Chuang, \textsl{Quantum Computation and Quantum Information} (Cambridge University Press, Cambridge, 2000).
\bibitem{GHZ1990} D. M. Greenberger, M. A. Horne, A. Shimony, and A. Zeilinger, ``Bell's theorem without inequalities,'' Am. J. Phys. \textbf{58}, 1131 (1990).
\bibitem{W2000} W. D$\ddot{\textrm{u}}$r, G. Vidal, and J. I. Cirac, ``Three qubits can be entangled in two inequivalent ways,'' Phys. Rev. A \textbf{62}, 062314 (2000).
\bibitem{FD2011} F. Fr$\ddot{\text{o}}$wis and W. D$\ddot{\text{u}}$r, ``Stable macroscopic quantum superpositions,'' Phys. Rev. Lett. \textbf{106}, 110402 (2011).
\bibitem{Yan2011} F. L. Yan, T. Gao, and E. Chitambar, ``Two local observables are sufficient to characterize maximally entangled states of $N$ qubits,'' Phys. Rev. A \textbf{83}, 022319 (2011).
\bibitem{GYE2014} T. Gao, F. L. Yan, and S. J.  van Enk, ``Permutationally invariant part of a density matrix and nonseparability of $N$-qubit states,'' Phys. Rev. Lett. \textbf{112}, 180501 (2014).
\bibitem{Kok2007} P. Kok, W. J. Munro, K. Nemoto, T. C. Ralph, J. P. Dowling, and G. J. Milburn, ``Linear optical quantum computing with photonic qubits,'' Rev. Mod. Phys. \textbf{79}, 135--174 (2007).
\bibitem{Pan2012} J. W. Pan, Z. B. Chen, C. Y. Lu, H. Weinfurter, A. Zeilinger, and M. $\dot{\text{Z}}$kowski, ``Multiphoton entanglement and interferometry,'' Rev. Mod. Phys. \textbf{84}, 777--838 (2012).
\bibitem{Milburn1984} G. J. Milburn and D. F. Walls, ``State reduction in quantum-counting quantum nondemolition measurements,'' Phys. Rev. A \textbf{30}, 56--60 (1984).
\bibitem{Imoto1985} N. Imoto, H. A. Haus, and Y. Yamamoto, ``Quantum nondemolition measurement of the photon number via the optical Kerr effect,'' Phys. Rev. A \textbf{32}, 2287--2292 (1985).
\bibitem{KLM2001} E. Knill, R. Laflamme, and G. J. Milburn, ``A scheme for efficient quantum computation with linear optics,'' Nature (London) \textbf{409}, 46--52 (2001).
\bibitem{SimonPan2002} C. Simon and J. W. Pan, ``Polarization entanglement purification using spatial entanglement,'' Phys. Rev. Lett. \textbf{89}, 257901 (2002).
\bibitem{ShengDeng2010} Y. B. Sheng and F. G. Deng, ``One-step deterministic polarization-entanglement purification using spatial entanglement,'' Phys. Rev. A \textbf{82}, 044305 (2010).
\bibitem{DYG2013} D. Ding, F. L. Yan, and T. Gao, ``Preparation of $km$-photon concatenated GHZ states for observing distinct quantum effects at macroscopic scale,'' J. Opt. Soc. Am. B, \textbf{30}, 3075--3078 (2013).
\bibitem{DingYanGao2012} D. Ding, F. L. Yan, and T. Gao, ``Entangler and analyzer for multiphoton maximally entangled states using weak nonlinearities,'' e-print arXiv:quant-ph/1209.6118 (2012).
\bibitem{Kwiat1995} P. G. Kwiat, K. Mattle, H. Weinfurter, A. Zeilinger, A. V. Sergienko, and Y. Shih, ``New high-intensity source of polarization-entangled photon pairs,'' Phys. Rev. Lett. \textbf{75}, 4337--4341 (1995).
\bibitem{GHZExperiment99} D. Bouwmeester, J. W. Pan, M. Daniell, H. Weinfurter, and  A. Zeilinger, ``Observation of three-photon Greenberger-Horne-Zeilinger entanglement,'' Phys. Rev. Lett. \textbf{82}, 1345--1349 (1999).
\bibitem{Pan2001} J. W. Pan, M. Daniell, S. Gasparoni, G. Weihs, and A. Zeilinger, ``Experimental demonstration of four-photon entanglement and high-fidelity teleportation,'' Phys. Rev. Lett. \textbf{86}, 4435--4438 (2001).
\bibitem{DuQiao2012} K. Du and C. F. Qiao, ``Scalable generation and characterization of a four-photon twelve-qubit hyperentangled state,'' J. Mod. Opt. \textbf{59}, 611--617 (2012).
\bibitem{DingYan2013} D. Ding and F. L. Yan, ``Efficient scheme for three-photon Greenberger-Horne-Zeilinger state generation,'' Phys. Lett. A  \textbf{377}, 1088--1094 (2013).
\bibitem{GLP1998} P. Grangier, J. A. Levenson, and J. P Poizat, ``Quantum non-demolition measurements in optics,'' Nature (London) \textbf{396}, 537--542 (1998).
\bibitem{MNBS2005} W. J. Munro, K. Nemoto, R. G. Beausoleil, and T. P. Spiller, ``High-efficiency quantum-nondemolition single-photon-number-resolving detector,'' Phys. Rev. A \textbf{71}, 033819 (2005).
\bibitem{NM2004} K. Nemoto and W. J. Munro, ``Nearly deterministic linear optical controlled-NOT gate,'' Phys. Rev. Lett. \textbf{93}, 250502 (2004).
\bibitem{Barrett2005} S. D. Barrett, P. Kok, K. Nemoto, R. G. Beausoleil, W. J. Munro, and T. P. Spiller, ``Symmetry analyzer for nondestructive Bell-state detection using weak nonlinearities,'' Phys. Rev. A \textbf{71}, 060302(R) (2005).
\bibitem{Kok2008} P. Kok, ``Effects of self-phase-modulation on weak nonlinear optical quantum gates,'' Phys. Rev. A \textbf{77}, 013808 (2008).
\bibitem{HRB2009} B. He, Y. H. Ren, and J. A. Bergou, ``Creation of high-quality long-distance entanglement with flexible resources,'' Phys. Rev. A \textbf{79}, 052323 (2009).
\bibitem{LHBR2009} Q. Lin, B. He, J. A. Bergou, and Y. H. Ren, ``Processing multiphoton states through operation on a single photon: Methods and applications,'' Phys. Rev. A \textbf{80}, 042311 (2009).
\end{thebibliography}
\end{document}